\documentclass[12pt]{iopart}
\usepackage{iopams}
\usepackage{graphicx}
\usepackage{bm}
\usepackage{citesort}

\begin{document}

\title[Magnetically controlled current flow in coupled-dot arrays]
{Magnetically controlled current flow in coupled-dot arrays}

\author{Panagiotis Drouvelis$^1$\footnote{Panagiotis.Drouvelis@tyndall.ie},
Giorgos Fagas$^1$\footnote{Georgios.Fagas@tyndall.ie}
and Peter Schmelcher$^{2,3}$\footnote{Peter.Schmelcher@pci.uni-heidelberg.de}
}
\address{$^1$ Tyndall National Institute, Lee Maltings, Prospect Row, Cork, Ireland}
\address{$^2$ Theoretische Chemie, Universit\"at Heidelberg, Im Neuenheimer Feld 229,
D-69120 Heidelberg, Germany}
\address{$^3$ Physikalisches Institut, Philosophenweg 12, Universit\"at Heidelberg,
D-69120 Heidelberg, Germany}


\begin{abstract}
Quantum transport through an open periodic array of up to five dots is investigated in
the presence of a magnetic field. The device spectrum exhibits clear features of the
band structure of the corresponding one-dimensional artificial crystal which evolves with
varying field. A significant magnetically controlled current flow is induced with
changes up to many orders of magnitude depending on temperature and material
parameters. Our results put forward a simple design for measuring with current technology
the magnetic subband formation of quasi one-dimensional Bloch electrons.
\end{abstract}
\pacs{73.23.Ad, 73.21.La, 73.20.At, 73.23.-b}


\section{Introduction}

Single quantum dots are the solid state analogue of an atom whereas the properties
of coupled-dots may resemble that of molecules.
Quantum transport through open quantum dots, being an equally intriguing as well
as extensively investigated topic~\cite{Alhassid00}, continues to provide new
insights in fundamental phenomena and fuels a wealth of nanoelectronic
applications
~\cite{Reimann02,Kouwenhoven01,Alhassid00,chang94,akis97,chan95,PRL01OSS,vavilov05}.
Arrays of coupled-dots may be considered as one-dimensional artificial
crystals with the dot repeating unit acting as the lattice basis.
If the coupling is strong enough the formed Bloch states yield an electronic structure
that uncovers many similarities with the subbands of quasi one-dimensional systems with
a much reduced reciprocal lattice vector. It is also well known that a uniform
magnetic field applied to Bloch electrons yields magnetic subbands with an overall
different spectrum~\cite{Zak1_64,Brown1_64}. Unlike the lack of any impact
in one dimension, in two dimensions these form the famous Hofstadter
butterfly~\cite{PRB76H}. The effect of confinement in two dimensional
ribbons has been studied in Ref.~\cite{Munoz05}. The question remains open as
to what extent there exists an observable magnetic effect for the intermediate
dimensionality,
as in the case of an array of open quantum dots. Moreover,
experimental evidence in the literature is scarce~\cite{PRL01ASK,PRB91GWW} and the
effect of magnetic subbands is hard to isolate in the common setup of lateral
semiconductor superlattices~\cite{EPL03LSR}. Hence, the prospect of measuring its
properties in a simple fashion is quite attractive.

In this study we consider small coupled-dot arrays that present distinct spectral
properties regulated via an applied magnetic field $B$.
The electron transport exhibits bright and dark windows reflecting
an electronic structure that is reminiscent of the energy bands of the corresponding
linear artificial crystal. This unique feature allows to explore
the $B$-dependence of the subbands of the quasi one-dimensional Bloch electrons.
With varying magnetic field, our calculations demonstrate qualitative (and quantitative)
changes of the bright and dark transport windows in the suggested array structure,
thus, yielding a direct signature of the magnetic subband formation in the
magnetoconductance.

Coupled-dot arrays may also be used as elements in magnetoresistive devices~\cite{APL02SOB}.
Such a design of either chaotic~\cite{PRL01OSS} or rectangular~\cite{Elhassan04} quantum
dots in alignment has been recently realized with a split-gate technique.
In particular, the experiments of Ref.~\cite{Elhassan04} showed a large magnetoresistance
at a field slightly greater than the magnetic field $B_c$ that corresponds to an electron
cyclotron radius equal to the size $W$ of the dot, i.e., $B_c = \hbar k_F / eW$ ($k_F$ is
the Fermi wavevector). Despite the possible role of Bragg reflection in a periodic
array, in these experiments it is still unclear how just Bragg reflection of electrons
would result in the non-exponential drop of the conductance measured as the number of the
dots increases. In fact, new conductance features such as enhanced reflection can
be expected for $B/B_c \approx 1$ purely from the induced changes in the
classical dynamics due to commensurability between the cyclotron radius and $W$.
Here, by excluding any other contribution to the magnetoconductance, we expose the
pure quantum mechanical effect of Bragg reflection of electrons propagating phase-coherently
across the dot array. Magnetic subbands form for any $B/B_c<1$, thereby,
suppressing the conductance due to Bragg reflection at the newly formed band edges. Hence,
as $B$ varies, the fingerprints of single bands are exposed via a significant
magnetoresistance even at moderate field strengths.

\begin{figure}[t]
\centering
\includegraphics[width=13.5cm]{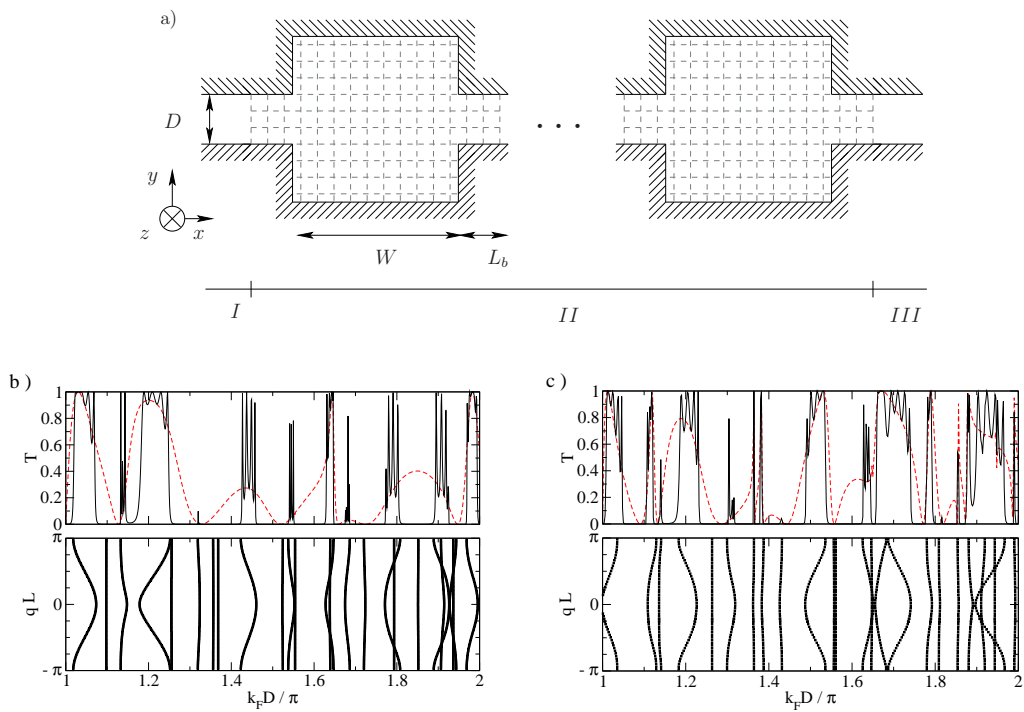}
\caption{(Color online)
(a) Schematic representation of the discussed open array of quantum dots.
(b) Upper panel: field-free quantum transmission through a single-dot (dashed curve)
and the five-dot array of (a). Lower panel: energy spectrum of the
corresponding one-dimensional artificial crystal with lattice spacing
$L = W + L_b$. Note that flat energy bands do not contribute to
transport since electrons acquire zero group velocity.
(c) Same as (b) for a magnetic flux $\Phi \approx 0.7 \phi_0$ piercing the unit
cell. We recall that the integer part of $k_FD/\pi$ indicates the number of propagating
channels in the leads and $q$ defines the Bloch vector of the periodic structure.
}
\label{figure1}
\end{figure}

\section{Formulation of the problem}

Fig.~\ref{figure1}(a) shows the setup in discussion.
We assume that square quantum dots of size $W$ are laterally confined on
the surface of a semiconductor heterostructure by an electrostatic field
which creates effective hard wall boundaries for ballistically propagating electrons.
The coupled leads are modelled by quasi one-dimensional conduction band
electrons freely propagating along the $x$ - direction with a Fermi distribution
$f_K(E)=[exp(\frac{E-\mu_K}{k_BT})+1]^{-1}$, $\mu_K = E_F \pm eV_{SD}/2$
being the chemical potential in the left (K=L) and right (K=R) lead when a bias
voltage $V_{SD}$ is applied. The point contacts bridging the dots have
square geometry of dimensions $L_b = D = 0.3 W$ that are of the
order of the Fermi wavelength $\lambda_F = 2\pi/k_F$. Although quantitative details differ,
our conclusions are independent of this simplest design.

We model the electronic structure via a single-band effective mass equation
of electrons in a magnetic field. The corresponding Hamiltonian is
$H = \frac{({\bf p} - e{\bf A})^2}{2 m^*} + U({\bf r})$
where $m^*$ is the effective mass (fixed to $0.05m_e$ unless otherwise stated).
The boundary conditions are imposed via the confining potential $U({\bf r})$, which is chosen to be zero
inside the enclosed area in Fig.~\ref{figure1}(a) and infinite outside.
The Hamiltonian is discretized on a two-dimensional tight-binding grid
using the Peirls' substitution for the vector potential ${\bf A}$
and can be expressed in the second quantization form~\cite{Ferry97}:
\begin{equation}
H({\mathbf{r}}) =
\sum\limits_{\mathbf{r}} \epsilon_{\bf{r}} c_{\bf{r}}^\dagger c_{\bf{r}} +
\sum\limits_{\bf{r},\bf{\Delta r}}
(Ve^{2 \pi i \frac{ {\bf{A(r)}} {\bf{\Delta r}} }{\phi_0} }
c_{\bf{r}}^\dagger c_{\bf{r}+\bf{\Delta r}} + h.c.).
\label{eq1}
\end{equation}
Here, the $c_{\bf r}^\dagger$ ($c_{\bf r}$) define a set of creation (annihilation)
operators on each site of the grid and
${\bf{\Delta r}}$ indicates the vector from the site {\bf{r}} to
its nearest neighbors. The on-site energy is $\epsilon_{\bf{r}} = 4V$ with the hopping matrix element
$V = \hbar^2 / 2m^*a^2$; $a$ is our lattice mesh
constant. The  magnetic field ${\bf B} = B{\bf z}$, which is applied in the region $II$
of Fig.~\ref{figure1}(a) and is zero in regions $I$ and $III$,
is introduced via the vector potential ${\bf{A}}$ in the Peierls phase factor;
$\phi_0 = h/e$ is the flux quantum.
Charge transport properties are calculated within the Landauer scattering
approach which expresses the current as follows:
\begin{equation}
I = \frac{2e}{h} \int\limits_{-\infty}^{\infty} {\rm T}(E) (f_L(E) - f_R(E)) dE
\label{eq2}
\end{equation}
in which ${\rm T}(E)$ is the quantum transmission function for injected electrons with
energy E; the factor two accounts for spin degeneracy.
We calculate ${\rm T}$ using our parallel algorithm of the
recursive Green's functions method~\cite{Drouvelis06}. As the system
size increases one needs to invert a block-tridiagonal matrix which
scales linearly with the array length. For serial processing
this yields an additional cost that we avoid by distributing 
the scatterer's domain over several processors.

\section{Results}

The upper panels of Figs.~\ref{figure1}(b) and \ref{figure1}(c)
show the field-free and $B \ne 0$, respectively, quantum transmission in the first
open channel. Transport through a five-dot array is indicated by the solid curves.
In contrast to the single-dot transmission spectrum - plotted in dashed line -
bright and dark windows are formed in which transport is either allowed or suppressed.
These compare well with the energy bands and gaps of the electronic structure
of the corresponding infinite linear artificial lattice obtained independently
by direct diagonalization; this is shown for zero and finite $B$ in the lower panels of
Figs.~\ref{figure1}(b) and \ref{figure1}(c), respectively.
Also evident in those figures is the prominent dependence of the band structure
with respect to the magnetic flux piercing the unit cell.
Broad energy bands contribute electron states that are almost fully transmitted, whereas,
narrow sections exhibit weaker transmission signals.
The remarkable characteristic is that such a transmission spectrum is rapidly
obtained for a quantum dot array with just a few unit cells as can
be seen from the comparison of the upper and lower panels of Figs.~\ref{figure1}(b)
and \ref{figure1}(c). In practice, this facilitates the realization
of a device at length scales comparable to the electronic phase coherence length
at finite temperatures so that quantum features do not wash out.
It may seem natural that only a small number of dots can reproduce the signatures of the
artificial crystal, which is fully consistent with the fast convergence of the transmission
with length in similar systems for $B=0$ (see for example Ref.~\cite{Fagas04}).
It is less straightforward to account for the interplay of magnetic flux with lattice
periodicity for finite fields but the evident correlation of energy and transmission
spectra strongly supports the above physical interpretation of our numerical results.
\vspace*{0.5cm}
\begin{figure}[h]
\centering
\includegraphics[height=4.75cm,width=7.25cm]{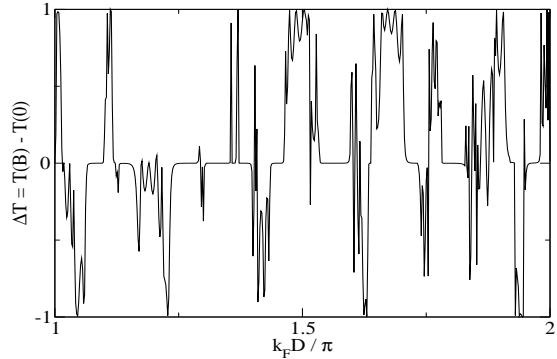}
\caption{
Magnetically controlled flow demonstrated via the profile of the difference of the quantum
transmission for the field-free and $B = 0.3B_c$ cases. }
\label{figure2}
\end{figure}

In Fig.~\ref{figure2}, we plot the transmission function difference
between the field-free structure and that at a field of strength
$B = 0.3B_c$. The positive and negative parts reflect the newly formed
magnetic subband structure of Bloch electrons in the corresponding
one-dimensional artificial crystal which cause the bright and dark
transport windows to occur at different spectral positions.
As discussed later, for a given geometry and Fermi energy (i.e., fixed $k_FD/\pi$)
the contrast in current flow due the differing transmission spectra can also be traced as
a function of magnetic field to yield the evolution of the magnetic subbands.
We note that there exists broad energy ranges over which bright transport
windows at non-zero magnetic field overlap with dark areas at vanishing $B$, e.g.,
at $k_FD/\pi \approx 1.5$ and $k_FD/\pi \approx 1.67$.
This feature marks a mechanism for magnetically controlled
current flow which can be realized up to the order of liquid nitrogen temperatures, as will be
shown below.

\begin{table}
\caption{
SI units at $k_FD/\pi = 1.5$ assuming $m^*=0.05m_e$.
}
\label{table1}
\centering
\begin{tabular*}{6cm}{@{\extracolsep{\fill}}cccc}
\\
$\lambda_F$(nm) & $W$(nm) & $E_F$(meV) & $B_c$(T) \\
 \hline
20 & 50 & 74.5 & 3.78\\
30 & 74 & 33 & 1.68 \\
50 & 123 & 11.9 & 0.6 \\
 \hline
 \hline
\end{tabular*}
\label{tab1}
\end{table}

Although the system of natural units is practically convenient when estimating the
upper limit of the conductance as determined by the number of open channels $k_FD/\pi$,
which is independent of the number of magnetic subbands that may occur in the dot array,
at this point it is instructive to interpret it to SI units.
Assuming $\lambda_F = 30$nm with $m^*=0.05m_e$ yields $E_F = 33$meV and $B_c = 1.68$T.
Regarding dimensions each quantum dot should be $W \approx 75$nm wide and the width of the
lead $D \approx 22$nm at $k_FD/\pi \approx 1.5$. The lattice spacing $L$ is around $100$nm
defining a total array length of less than $500$nm for five coupled-dots.
In a strict sense, these dimensions define the range of validity of our results
regarding temperature. Apart from the thermal broadening, the temperature controls
the scattering mechanisms determining the electronic phase coherence length. Since
we have so far assumed that electrons are coherently propagating, the array length
must be shorter than the latter~\cite{footnote1}.
More examples are presented in Table~\ref{tab1}. These show the interplay between
linear dimensions and $B_c$.
\vspace*{0.5cm}
\begin{figure}[t]
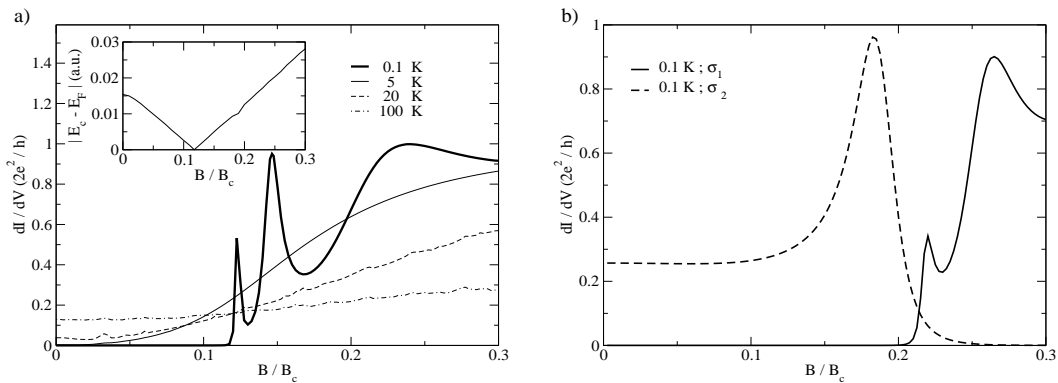

\centering
\includegraphics[height=5cm, width=6.7cm]{fig3a.eps} \hspace*{0.3cm}
\includegraphics[height=5cm, width=6.7cm]{fig3b.eps}
\caption{ (a) Linear-response magnetoconductance at various temperatures
($k_FD/\pi \approx 1.5$). Inset: magnetic field dependence of the
distance between the Fermi energy $E_F=74.5$meV and the band edge $E_c$
accounting for the resonant structure of the low-temperature
magnetoconductance when crossing occurs at
$B \approx 0.12 B_c \approx 0.45$T. (b) Typical magnetoconductance modified by the
presence of weak boundary disorder with relative strength
$\sigma_1 = 0.027 \langle W \rangle$ and $\sigma_2 = 0.061 \langle W \rangle$. }
\label{figure3}
\end{figure}

In Fig.~\ref{figure3}(a), we furnish our observations with the linear-response
magnetoconductance curve at various temperatures.
An overall increase of the conductance with magnetic field strength is clearly observed.
A remarkable feature, central to this work, is the fine peak-structure
of the magnetoconductance $dI/dV$ at very low temperatures which
relates to the formation of the spectrum of Bloch electrons in a magnetic
field. This is demonstrated in the inset of Fig.~\ref{figure3}(a).
As the band structure modifies with the magnetic field, the edge of a single
band $E_c$ crosses the Fermi energy $E_F$ at $B/B_c \approx 0.12$.
When the distance $\left|E_c-E_F\right|$ vanishes a bright transport
window is induced that gives rise to the resonant structure of 
$dI/dV$ in the sub-Kelvin regime (thick line in Fig.~\ref{figure3}(a)).
Due to the well-pronounced peaks one could think of using these
as a probe for the magnetic subband structure.
At higher temperatures thermal broadening causes averaging over a larger part of the spectrum
including many adjacent minibands and gaps. This increases the low-field
conductance whereas simultaneously decreases the corresponding higher field
values.

Below we would like to discuss some aspects regarding the effects of disorder in our samples.
Realistic devices may be experimentally prepared with electronic mean free paths
of the order of several $\mu m$ for the regime of lower temperatures
discussed herein \cite{Eroms05}. These clearly allow for ballistic transport experiments.
However, to account for a possible residual departure from the ideal
periodicity of our structure we investigate two samples whose boundaries along the current direction are
roughened with a standard deviation $\sigma$; in this case $\lambda_F$ is much larger than
the disorder's characteristic length scale. The magnetoconductance is shown in
Fig.~\ref{figure3}(b). For $\sigma_2 = 0.061 \langle W \rangle$
the disorder-potential imposes a fluctuation to the electron's energy,
of $\sim 2 \% E_F$ in the region of the square dots and $\sim 30 \% E_F$ in the
interdot bridge sections. The latter order of magnitude is consistent with the background
disorder observed in the experimental setup of Ref.~\cite{Elhassan04}.
The presence of the disorder considered here
modifies transport through the coupled-dot system in a two-fold way.
First, the disorder within the square dots' sections perturbes the position of the
bands such that they may coincide with $E_F$, as it is the case in the magnetoconductance curve corresponding
to the sample with $\sigma_2$. Second, the enhanced residual disorder at the bridge sections
weakens the coupling between the dots, thereby reducing the volume of the transport windows.
Effectively, this fact decreases the bands' width and respectively the amount of the transmitted current.
However, the overall resonant structure of the
magnetoconductance curve is preserved.
\vspace*{0.5cm}
\begin{figure}[h]
\centering
\includegraphics[height = 5.2cm, width=7.2cm]{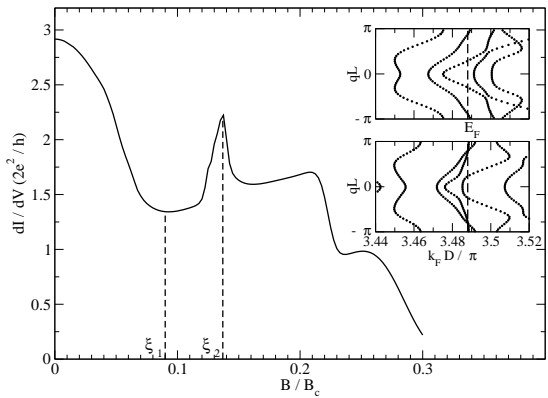}
\caption{
Linear-response magnetoconductance at $T = 0.1 K$ and $k_FD/\pi \approx 3.5$.
Insets depict the band structure close to the Fermi energy at two selected
magnetic field strengths $\xi_1 = 0.9$ (upper) and $\xi_2 \approx 0.14$ (lower)
corresponding to two distinct conduction regimes with $1< dI/dV <2$ and $2< dI/dV <3$,
respectively. The dashed vertical lines indicate the Fermi energy.
}
\label{figure4}
\end{figure}

Fig.~\ref{figure4} shows the conductance profile at $E_F$ corresponding to a higher number
of propagating modes in the leads yielding an upper limit of three.
The magnetoconductance reveals a much richer structure. However, it is uniquely
distinguished into regions where the maximum attainable value of the conductance differs
depending on the number of the dot-array bands that the incoming electrons
can populate. As illustrated using the insets in Fig.~\ref{figure3}, with varying magnetic field
more magnetic subbands are formed at the fixed $E_F$ which can be further occupied by
the incoming electrons. As in Fig.~\ref{figure3}, this effect leads to a pronounced peak marking
the magnetoconductance yielding again information about the magnetic
subband structure at higher energies.

\begin{figure}[t]
\centering
\includegraphics[angle=-90,width=7.5cm]{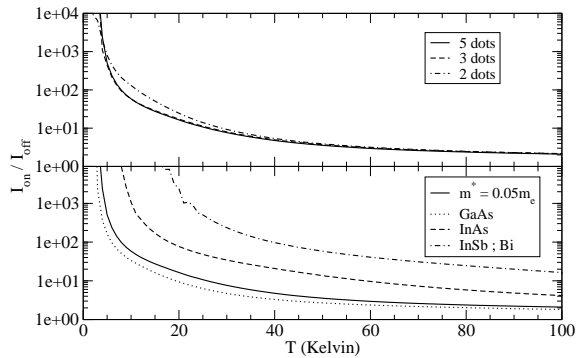}
\caption{
(Upper panel) Ratio $I_{on}/I_{off}$ of the current flow in the on
($B = 0.3B_c \approx  1.13$T) and off ($B = 0$)
state as a function of temperature
for an array of $N = 2,3,5$ coupled-dots.
(Lower panel) Temperature dependence of the $I_{on}/I_{off}$ ratio
for various effective masses $m^*$.
$k_FD/\pi$ and $E_F$ are the same as in Fig.~\ref{figure3}.
}
\label{figure5}
\end{figure}

For fixed $k_FD/\pi$, it is interesting to analyse the effect of temperature and
variations to the effective mass as exposed by various materials. To this end,
we define the enhancement (on) - suppression (off) ratio of current
flow $I_{on}/I_{off}$ in the linear response regime when we apply the
highest magnetic field we considered, i.e., $B = 0.3B_c$.
In the upper panel of Fig.~\ref{figure5}, the temperature dependence of the
$I_{on}/I_{off}$ ratio is shown for an array with varying number $N$ of coupled-dots.
Remarkably enough the results hardly modify with $N \gtrsim 3$ in support of our previous
remarks. We observe that relatively large ratios in excess of $100$ can be achieved for
temperatures up to $\sim 10$K and can be preserved to $I_{on}/I_{off} > 10$ for temperatures
up to $\sim 26$K. Note that this behavior may be drastically improved
with a selective choice of materials and geometry. A search in the parameter
space for the latter is best done with an exhaustive analysis of
the calculated magnetic subband structure at each field value which is beyond the
scope of this study. Rather, in the lower panel of Fig.~\ref{figure5} we show
that probing of the magnetic subband structure in materials with lighter effective mass
would be greatly enhanced and feasible at higher temperatures.

\vspace*{0.5cm}

\section{Concluding remarks}

A few additional comments regarding the effects of electronic interactions and spin
are in order. An estimation of the ratio of the single-electron charging energy $U$ for each
decoupled dot over the interdot coupling $t$ (determined mainly by $D / W$)
yields $U/t<1$ for most dot-modes with wide energy bands
forming in the array. In this case, there are small interaction corrections to the
conductance spectrum within the first conduction channel ($k_FD/\pi<2$) whose main effect is
to sharpen the resonances observed in the magnetoconductance (Fig.~\ref{figure3}), while
not modifying the main transport mechanism~\cite{Aleiner02}. At higher $k_FD/\pi$
interaction effects are even less significant. Interactions may modify the transport
profile for some distinct very narrow bands at $k_FD/\pi<2$ for which $U/t>1$ or if
$D/W$ is greatly reduced. Such an analysis of weakly coupled dots is not within
the scope of the present study. Finally, the addition of the Zeeman splitting
due to spin provides very small corrections to the energy and can be safely neglected for
the magnetic fields considered here.

To summarize, we have presented an investigation of ballistic transport through
a finite array of coupled-dots from the perspective of a quantum
mechanical magnetically tunable mechanism that redefines bright and
dark transport windows. The latter have been respectively identified as the energy
bands and gaps of the electronic structure of the corresponding one-dimensional
artificial crystal despite the small number of dots. Thus, by
tracing their magnetic field dependence we showed that the precursor of magnetic subband
formation in the energy spectrum can be readily observed.
The broad energy range of the transport windows also reveals a well defined mechanism
that yields magnetically controlled currents with large enhancement - suppression ratios
which can extend up to several tens of Kelvin depending on material parameters.
With present technology such a device can be realized within a
region of $\sim 300$ nm at a magnetic field of $\sim 0.5$T.

\ack
The authors are grateful to J.~Eroms and I.~Knezevic for their critical reading of 
the manuscript and fruitful comments.
P.\,D. acknowledges financial support from the Deutsche Forschungsgemeinschaft in
the framework of the International Graduiertenkolleg IGK 710
and the Irish Research Council for Science, Engineering and Technology.
G.\,F. is thankful to the Science Foundation Ireland for funding.
P.\,S. gratefully acknowledges financial support from the Deutsche Forschungsgemeinschaft.


\begin{thebibliography}{22}
\bibitem{footnote1}
In the ballistic regime the phase-relaxation length may be estimated according to Ref.~\cite{Giuliani82}.
For an electron density corresponding to $E_F = 33$ meV this is $\sim 900$ nm at $T = 40$K and
reduces to $\sim 170$ nm at $T = 100$ K, which is few times smaller than the length of our device.
\end{thebibliography}
\end{document}